\documentclass[smallextended]{svjour3}       

\smartqed  

\usepackage{graphicx}
\usepackage{amsmath}
\usepackage[latin1]{inputenc}
\usepackage{txfonts}
\usepackage{hyperref}
\usepackage{float}
\usepackage{xcolor}
\usepackage[noadjust]{cite}
\usepackage{rotating}

\newcommand{\be}{\begin{equation}}
\newcommand{\ee}{\end{equation}}
\newcommand{\R}{\mathbb{R}}
\newcommand{\xp}{R_t}
\newcommand{\xm}{L_t}

\journalname{Journal of Statistical Physics}

\begin{document}

\title{From weakly chaotic dynamics to deterministic subdiffusion via copula modeling}
\author{Pierre Naz\'e}
\institute{Pierre Naz\'e \at
             Unicamp, Instituto de F\'isica ``Gleb Wataghin'', DFCM, 09210-170, Campinas, SP, Brazil \\
             \email{p.naze@ifi.unicamp.br} \\
}

\date{}

\maketitle

\begin{abstract}

Copula modeling consists in finding a probabilistic distribution, called copula, whereby its coupling with the marginal distributions of a set of random variables produces their joint distribution. The present work aims to use this technique to connect the statistical distributions of weakly chaotic dynamics and deterministic subdiffusion. More precisely, we decompose the jumps distribution of Geisel-Thomae map into a bivariate one and determine the marginal and copula distributions respectively by infinite ergodic theory and statistical inference techniques. We verify therefore that the characteristic tail distribution of subdiffusion is an extreme value copula coupling Mittag-Leffler distributions. We also present a method to calculate the exact copula and joint distributions in the case where weakly chaotic dynamics and deterministic subdiffusion statistical distributions are already known. Numerical simulations and consistency with the dynamical aspects of the map support our results.

\end{abstract}

\section{Introduction}
\label{sec:1} 

Transport phenomena behaving differently from the usual Brownian motion have been detected by several experiments over the last century \cite{richardson1927, bohm1949, koenig1972, bernasconi1979, cardoso1988, solomon1993, dieterich2008, abe2015}, in particular a type of subdiffusive process, where the mean squared displacement grows proportional to $t^\alpha$, with $0<\alpha<1$. This relatively simple characteristic made it an object of intense research and theoretical models were created trying to describe it as well \cite{metzler2000, klages2008}. In this work, we study this phenomenon using Geisel-Thomae map \cite{geisel1984}, a dynamical system that reproduces a subdiffusive process when one observes the trajectory of a set of initial conditions. For being a spatially extended Pomeau-Manneville map \cite{pomeau1980}, Geisel-Thomae map preserves the weakly chaotic dynamics characteristic of that system \cite{akimoto2010}, being therefore a deterministic subdiffusion model with this specific dynamical behavior. Techniques such as continuous-time random walk (CTRW) \cite{montroll1965, zumofen1993} and infinite ergodic theory \cite{akimoto2010} look carefully into the relation between these phenomena, coming to a conclusion that the subdiffusive process observed in such system is a straightforward consequence of its weakly chaotic dynamics. In this work, we deepen our knowledge about their relation, connecting both phenomena by their characteristic statistical distribution. In order to do so, we present the technique of copula modeling.

Two situations are possible in a study of statistical dependency across random variables: the random variables can be statistically independent of each other or not. The former case is characterized by the joint cumulative distribution function (CDF) being given by the product of the variables CDFs. The latter case, on the other hand, presents a non-trivial functional form, in which finding its joint CDF becomes a hard task to be accomplished given that there is no specific model to be achieved in principle. One way out to solve that problem is the fundamental Sklar theorem \cite{sklar1959}, which says that, given two random variables $X$ and $Y$, their joint CDF $J$ can be expressed, in a unique way, as 
\be
J(x,y) = C(F_X(x), F_Y(y)),
\label{eq:copulapn}
\ee 
where $F_X$ and $F_Y$ are the respective marginal CDFs of $X$ and $Y$, and $C$ is a copula, a joint CDF defined on the unit sized square with additional properties (see \cite{nelsen2006} for more details). Copula modeling consists then in dividing the statistical dependency of random variables into marginal and copula distributions and using statistical methods to infer them properly. Examples of copula distributions are very well known in the literature (see \cite{nelsen2006, durante2010, trivedi2007} and references therein) and computationally accessible by copula modeling software package \cite{rdoc-copula}.

Here the central idea of this work. Consider $X_t$ as the random variable of jumps executed by the particle until a time $t$, according to Geisel-Thomae map. Rewriting $X_t=R_t-L_t$, where $R_t$ and $L_t$ are respectively the sum of the jumps done only to the right and left senses, we express the jumps distribution associated to $X_t$ as an expression of the joint distribution of $R_t$ and $L_t$. We perform then a copula modeling, where these new random variables will work as the marginal distributions. Thus, as $R_t$ and $L_t$ are connected to the number of first-passages of Pomeau-Manneville maps, whose distribution characterize the weakly chaotic dynamics \cite{gaspard1988, aizawa2007, venegeroles2012, naze2014}, they will obey the same statistical quantity. Therefore, the jumps distribution of deterministic subdiffusion is connected to the distribution of weakly chaotic dynamics.
   
The article is organized as follows. In section \ref{sec:2} we present the Geisel-Thomae map, its subdiffusivity and the connection between the distributions of weakly chaotic dynamics and deterministic subdiffusion. In section \ref{sec:3}, the marginal distributions are calculated exactly by techniques of infinite ergodic theory and the copula one inferred by statistical methods. We present also a method to calculate exactly the copula and joint distribution of the system since the statistical distribution already mentioned are known. In section \ref{sec:4}, we summarize what we have done, emphasize to physics community the importance of copula modeling and discuss perspectives from this work.

\section{Anomalous diffusion via copula modeling}
\label{sec:2}

The Pomeau-Manneville map is a function $T$, defined on the unit interval, whose expression is given by
\be
T(x) = x+{(2x)}^{1+\frac{1}{\alpha}} \quad (\text{mod 1}),
\label{eq:pm}
\ee
with $\alpha>0$. The parameter $\alpha$ determines its dynamics: if $\alpha\ge 1$, Pomeau-Manneville map is chaotic; if $0<\alpha<1$, the system is weakly chaotic. In this last case, the trajectory of the particle passes through an intermittent regime, where it spends much time near the laminar region, located about the neutral point $x=0$, and eventually visits the turbulent one, located in the remaining part of the phase space. Because of this, the time evolution of dynamical observables occurs at a sublinear rate and its conventional time average approaches to zero for long times \cite{gaspard1988}. To capture some chaotic aspects of the system, the time average is modified using the transformation $t\rightarrow t^\alpha$ in its normalization constant, leading us to a new type of ergodicity, where this new time average obeys an universal non-atomic distribution for a large class of dynamical observables \cite{aaronson1997, zweimuller2000}.
\begin{figure}[!ht]
 \centering
 \includegraphics[scale=0.7]{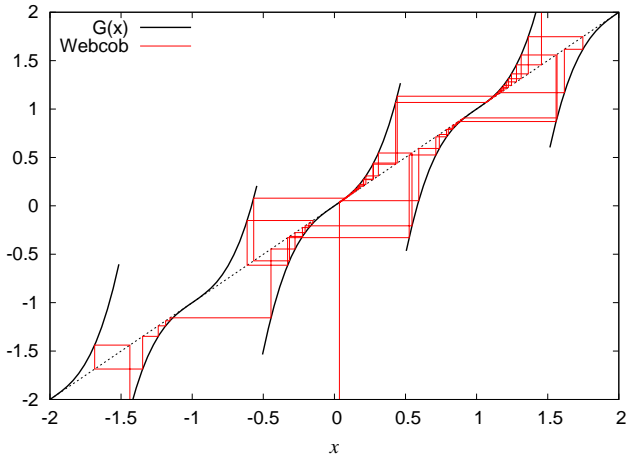}
 \caption{(Color online) Cobweb diagram of a Geisel-Thomae map $G$. The chain of maps was generated by the map $f(x)=x+{(2x)}^3$, being used an itinerary of $160$ jumps beginning from $x_{0}=0.035$. The dotted, thicker solid black and thinner solid red lines respectively represent the identity function, Geisel-Thomae map and the webcob. }
\label{fig:chain}
\end{figure}

To construct Geisel-Thomae map, we consider initially a Pomeau-Manneville map $T$, with $0<\alpha<1$, defined on the interval $[0,1/2]$. We start creating an odd function $M$, defined on $(-1/2,1/2]$, in such a way that $M(x)=T(x)$, for $x\in[0,1/2]$, and $M(x) = -T(-x)$, for $x\in(-1/2,0)$. Using then the property of displacement of degree one, $G(x+N):=M(x)+N$, where $N$ is a integer number, we construct Geisel-Thomae map $G$, now defined all over the real line. Summarizing such steps, one has
\be
G(x) := \begin{cases} x+{[2(x-N)]}^{1+\frac{1}{\alpha}},\quad x\in[N,N+1/2] \\ 
                      x-{[2(N-x)]}^{1+\frac{1}{\alpha}},\quad x\in(N-1/2,N] \end{cases},
\label{eq:chaing}
\ee
for all $N$ integer. The FIG. \ref{fig:chain} depicts a cobweb diagram of $G$.

During the computational simulation of this model we consider a sample of $n$ particles distributed uniformly over the interval $(-1/2,1/2)$, such that each one of them is iterated $t$ times by Geisel-Thomae map. The set of the $n$ final points generated by this procedure yields to a sample variance, which grows as a power-law with exponent $0<\alpha<1$, indicating that the model presents subdiffusion \cite{geisel1984}. This aspect is also revealed in the study of the PDF $\rho(x,t)$ of the particle position $x$ at time instant $t$, in which techniques from CTRW \cite{zumofen1993} or infinite ergodic theory \cite{miyaguchi2013} have shown that such PDF is the solution of time-fractional diffusion equation \cite{mainardi2001}, given by
\be
\rho_{\alpha}(x,t) = \frac{1}{\sqrt{4Dt^\alpha}}M_{\frac{\alpha}{2}}\left(\frac{|x|}{\sqrt{4Dt^\alpha}}\right),
\label{eq:rrrrho}
\ee
where $M_{\nu}$ is the Mainardi function, given in turn by
\be
M_{\nu}(z) = \frac{1}{\pi}\sum_{n=0}^\infty\frac{{(-z)}^n}{n!\Gamma[-n\nu+(1-\nu)]},
\label{eq:mainardi}
\ee
where $0<\nu<1$, and $D$ is the diffusion constant.

\begin{figure}
 \centering
 \includegraphics[scale=0.46]{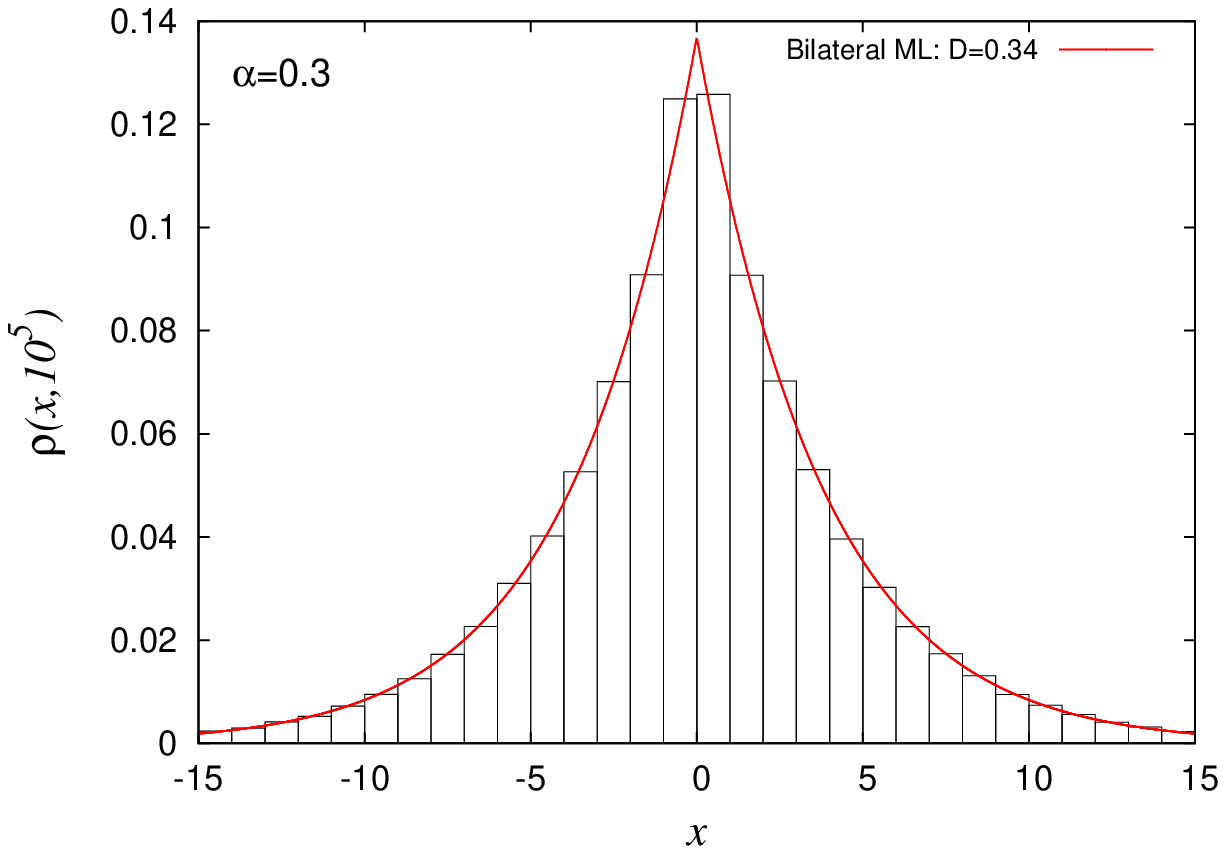}
 \includegraphics[scale=0.46]{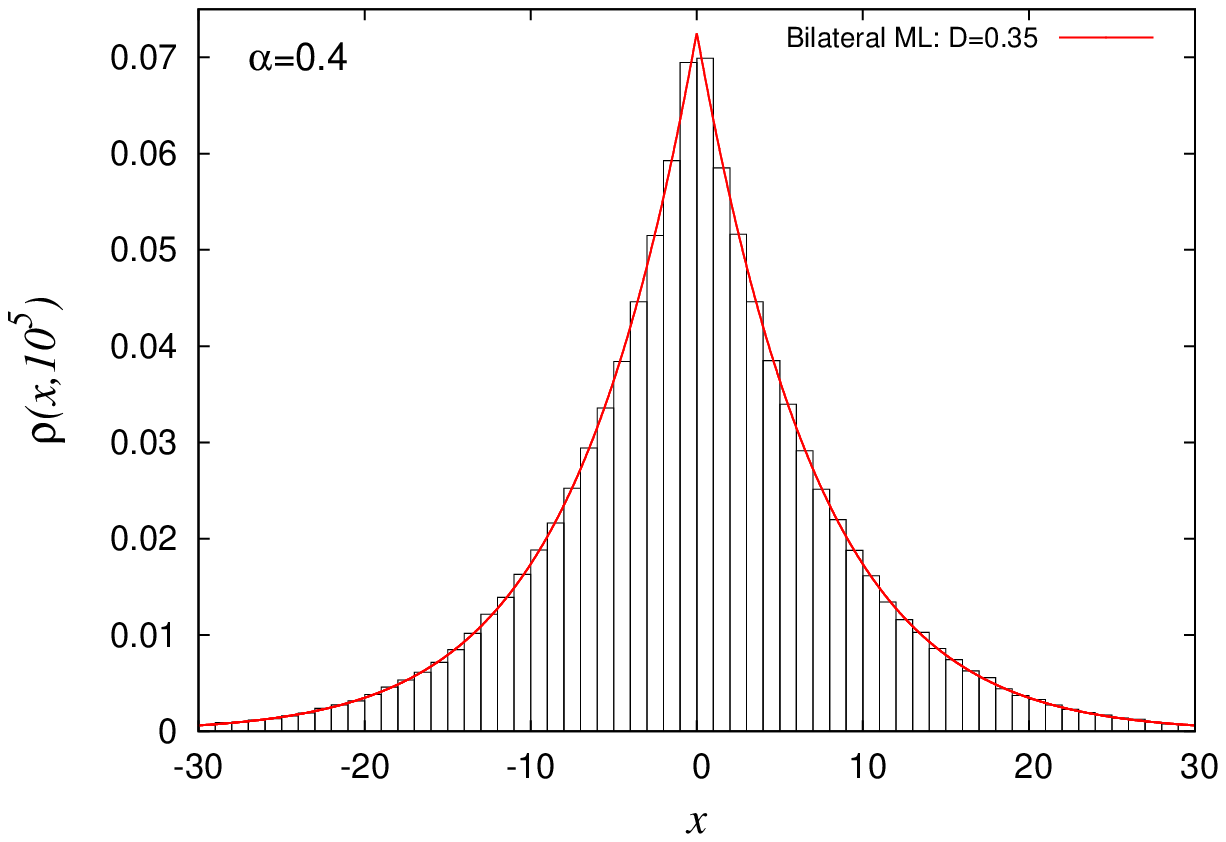}
 \includegraphics[scale=0.46]{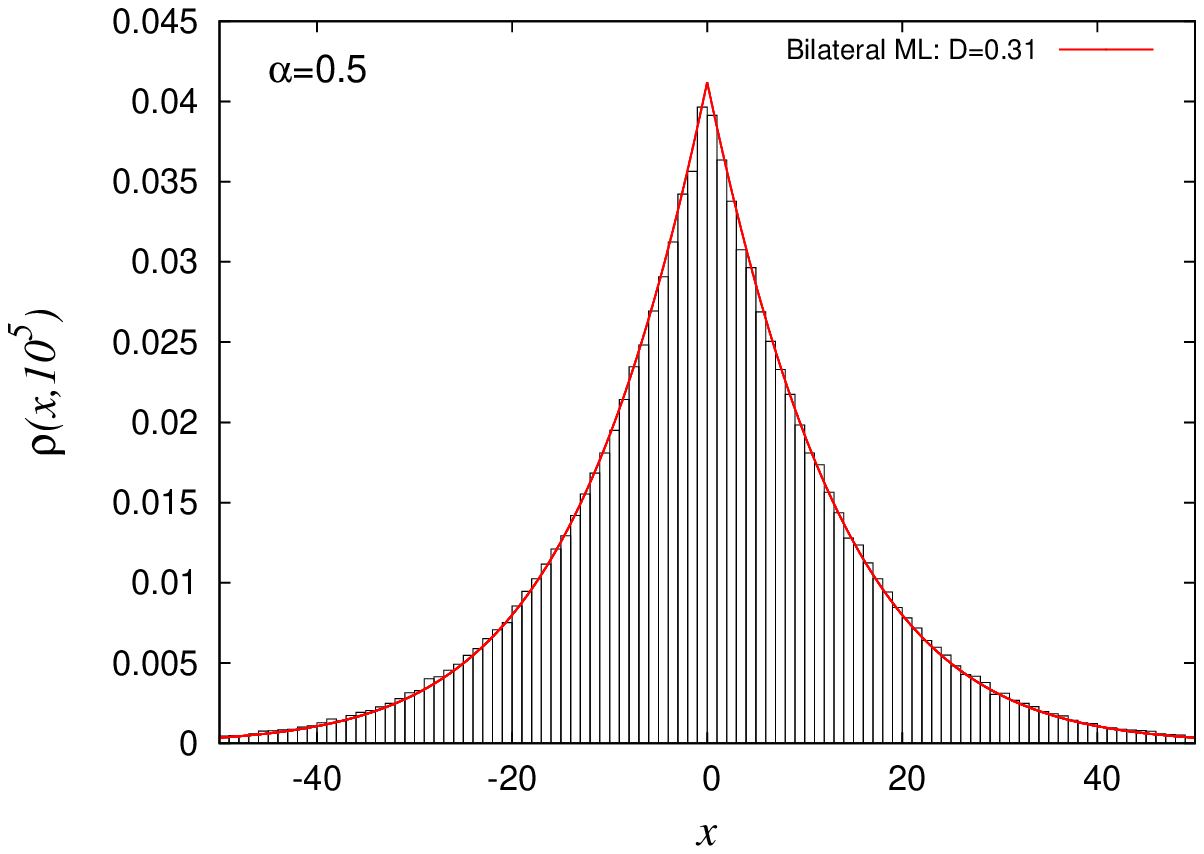}
 \includegraphics[scale=0.46]{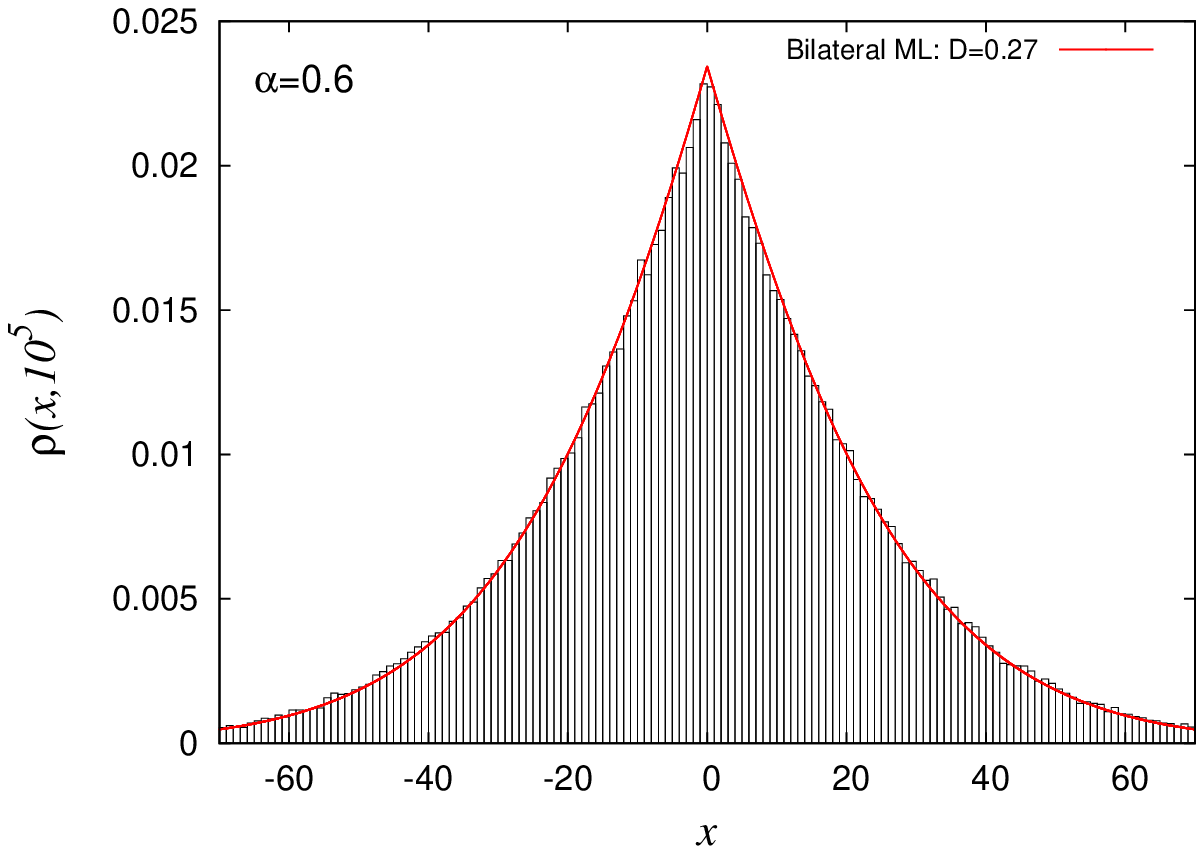}
\caption{(Color online) Comparisons between the PDF outlines generated by Eq. (\ref{eq:rrrrho}) (red lines) with the histogram produced by Geisel-Thomae map for $\alpha=0.3, 0.4, 0.5, 0.6$. The diffusion constants was chosen by inspection and the number of initial conditions was $n=250000$. For all cases, the agreement is very good.}
\label{fig:comp_rho}
\end{figure}

Thus, to show to the reader that Geisel-Thomae map exhibits subdiffusion, Fig. \ref{fig:comp_rho} shows comparisons of the PDF outlines generated by Eq. (\ref{eq:rrrrho}) with the histograms of final position $x$ at time $t$ for different parameters $\alpha$s. The agreement is very good indeed. Finally, for the purposes of outlines comparisons, we remark that the diffusion constants were taken by inspection, not choosing any particular theoretical model to estimate it.

To connect the statistical distributions of weakly chaotic dynamics and deterministic subdiffusion distribution, we consider the general situation where the random variable $X_t$ of the jumps executed until a time $t$ is rewritten as $X_t=R_t-L_t$, where $R_t$ and $L_t$ are respectively the jumps done only to the right and left senses. In this manner, the probability distribution function (PDF) $\rho(x,t)$ of $X_t$ is related to the joint PDF $j_t$ of $R_t$ and $L_t$ in the following way
\be
\rho(x,t) = \int_{0}^{\infty}j_t(|x|+z,z)dz.
\label{eq:diff}
\ee
Substituting the copula $C_t$ and the marginal CDFs $F_{R_t}$ and $F_{L_t}$ in Eq. ($\ref{eq:copulapn}$), and such result in its PDF version in Eq. ($\ref{eq:diff}$), one has
\be
\rho(x,t) = \int_{0}^{\infty}c_t(F_{\xp}(|x|+z),F_{\xm}(z))\rho_{\xp}(|x|+z)\rho_{\xm}(z)dz,
\label{eq:difffinal}
\ee
in which $c_t$ is the PDF of the copula $C_t$ and $\rho_{\xp}$, $\rho_{\xm}$ are the PDFs associated to the marginal distributions. Eq. (\ref{eq:difffinal}) is the main result of this work: applied to Geisel-Thomae map, it connects the statistical distributions of deterministic subdiffusion and weakly chaotic dynamics, as it will be demonstrated in the next section when the marginal distributions will be calculated. 

We remark one more time that our approach is completely general. If one is studying a system -- which does not need to be a map -- that produces any kind of anomalous diffusion, and whose PDF of jumps distribution is unknown, the copula modeling can be applied in the same manner. The advantage of Geisel-Thomae map, as we are going to see, is to find the marginal distribution exactly. Even if it is not possible find it exactly, methods of statistical inference could be applied to find approximations of the marginal distributions \cite{degroot2011}.

\section{Results}
\label{sec:3}

\subsection{Marginal distributions}

Given Geisel-Thomae map $G$ and a time of iteration $t$, consider two random variables, $\xp$ and $\xm$, such that the former is the absolute value of the sum of the displacements of jumps executed to the positive sense and the latter to the negative one:
\be
\xp := \sum_{k=0}^{t-1} \vartheta_{R}(G^k(X_{0})), \quad \xm := \sum_{k=0}^{t-1} \vartheta_{L}(G^k(X_{0})),
\ee
where the following observables are defined as
\be
\vartheta_{R}(x):=(G(x)-x)H(G(x)-x), 
\label{eq:vartheta}
\ee
and
\be
\vartheta_{L}(x):=(x-G(x))H(x-G(x)), 
\label{eq:vartheta}
\ee
being $H$ the step Heaviside function, $G^k$ the $k$-th iteration of $G$ and $X_0$ the uniform distribution defined over $[-1/2,1/2]$. Using the Eq. (\ref{eq:chaing}) in Eq. (\ref{eq:vartheta}), one has
\be
\vartheta_{R}(x) = \begin{cases} {[2(x-N)]}^{1+\frac{1}{\alpha}}&,\quad x\in[N,N+1/2] \\ 
                      0&,\quad  x\in(N-1/2,N] \end{cases},
\label{eq:varthetap}
\ee
and
\be
\vartheta_{L}(x) = \begin{cases} 0 &,\quad x\in[N,N+1/2] \\ 
                      {[2(N-x)]}^{1+\frac{1}{\alpha}}&,\quad x\in(N-1/2,N] \end{cases}.
\label{eq:varthetan}
\ee

We first remark that the marginal distribution must be identical for $\xp$ and $\xm$. The initial conditions uniformly distributed over $[-1/2,1/2]$ produce a symmetry whereby, if we have a particle in the position $x_{0}$, we must have another in $-x_0$. Moreover, by the map symmetry, the position of the particle must be such that $G^t(x_{0})=-G^t(-x_{0})$ at any time $t$. Thus the displacements will be different only by the sign, which will be vanished by the absolute values of the observables $\vartheta_{R}$ and $\vartheta_{L}$. Therefore if the PDFs of $\xp$ and $\xm$ are respectively $\rho_{\xp}$ and $\rho_{\xm}$, we must have  
\be
\rho_{\xp}(x,t) = \rho_{\xm}(x,t);
\ee
for now on we are only referring to $\rho_{\xp}$.

Before proceeding in our analysis, we briefly discuss the Aaronson-Darling-Kac (ADK) theorem, a fundamental result for determining the marginal distributions of Geisel-Thomae map. Consider a map $T$ conservative, ergodic, measure-preserving transformation on its phase space $A$ and $\mu$ is its invariant measure. The ADK theorem says that a suitable time average of an observable $\vartheta\in\mathcal{L}_{+}^{1}(\mu)$ converges in distribution to a random variable $\xi_{\alpha}$, which is scaled by the ensemble mean of the same variable. That is 
\be
\frac{1}{a_t}\sum_{k=0}^{t-1}\vartheta(T^k(X)) \overset{d}{\rightarrow} \xi_{\alpha}\int_{A}\vartheta d\mu,
\label{eq:mltheo}
\ee
where $(a_t)|_{t=0}^{\infty}$ is the return sequence, $X$ a random variable and $\xi_\alpha$ the normalized Mittag-Leffler distribution of order $\alpha$, with $0<\alpha<1$. For more information, one can see \cite{aaronson1997, zweimuller2000, aizawa2007, akimoto2010, venegeroles2012, naze2014}.

Returning to our discussion, the similarity between Geisel-Thomae map $G$ and the Pomeau-Manneville map $T$ suggests that the former is also an weakly chaotic map and it would be appropriate to use ADK theorem in the observable $\vartheta_{R}$ to determine $\rho_{R_t}$. However, we cannot proceed with such idea, because the domain of $G$ is the real line, which is not covered by the usual theory \cite{aaronson1997}. To circumvent this aspect, Akimoto and Miyaguchi \cite{akimoto2010} have pointed out that $G$ could be reduced to an weakly chaotic map for the purpose we have in mind. Observable given in Eq. (\ref{eq:varthetap}) has the property of evaluating not from the value of $x$, but only from its difference with its nearest integer. In that case, if a particular map $\bar{G}$ produces the same differences for some observable $\bar{\vartheta_R}$ and obeys to the ADK theorem, so the same theorem is valid for $G$ and the observable $\vartheta_R$. Analyzing the structure of our chain, a possible way to use this fact is considering
\be
 \bar{G}(x) := \begin{cases} x+{(2x)}^{1+\frac{1}{\alpha}} \quad\quad\,\, , \quad x\in[0,1/2] \\ 
                      x-{[2(1-x)]}^{1+\frac{1}{\alpha}}\, , \quad x\in(1/2,1] \end{cases}\, \text{mod 1},
\label{eq:reducedg}
\ee
with initial conditions $\bar{x_0}\in[0,1]$ \footnote{The correspondence between the initial conditions from each map is: $x_0 = \bar{x_0}$, for $x_0\geq 0$ and $x_0 = \bar{x_0}-0.5$, for $x_0< 0$. Again, we note that the only thing important here is the difference between successive jumps.} and the observable
\be
\bar{{\vartheta}_R}(x) = \begin{cases} {(2x)}^{1+\frac{1}{\alpha}}&,\quad x\in[0,1/2] \\ 
                      {[2(1-x)]}^{1+\frac{1}{\alpha}}&,\quad  x\in(1/2,1] \end{cases}.
\label{eq:varthetap2}
\ee
We observe that Eq. (\ref{eq:reducedg}) is well defined for $\alpha\in \R$ \footnote{Note that $\bar{G}\left(\frac{1}{2}\right)=\frac{1}{2}<1$, which is necessary for having only two branches on $[0,1/2]$. The same reasoning is valid for the interval $(1/2,1]$.}. $\bar{G}$ is an weakly chaotic map with finite domain and obeys therefore the ADK theorem. Thus, the evaluation of $\rho_{R_t}$ will lead to the same results. Considering Eq. \ref{eq:mltheo} and using $\vartheta=\vartheta_{R}$ and $a_t\sim t^\alpha/\beta$ for large times, one has
\be
\xp := \sum_{k=0}^{t-1} \vartheta_{R}(G^k(X_{0})) \overset{d}{\sim} \frac{t^{\alpha}}{\beta}\xi_{\alpha},
\ee
which lead us to
\be
\rho_{\xp}(x) \sim \frac{\beta}{t^\alpha}\rho_{\xi_{\alpha}}\left(\frac{\beta x}{t^\alpha}\right),
\label{eq:rhopn}
\ee
for large $t$, in which $\rho_{\xi_{\alpha}}$ is the PDF of $\xi_\alpha$ and $\beta$ is a constant in time, in principle dependent on $\alpha$. In another words, the positive displacement $R_t$ obeys a Mittag-Leffler distribution which uniformizes itself as time grows up according to a power-law.
	 
\begin{figure}
  \centering
    \includegraphics[scale=0.7]{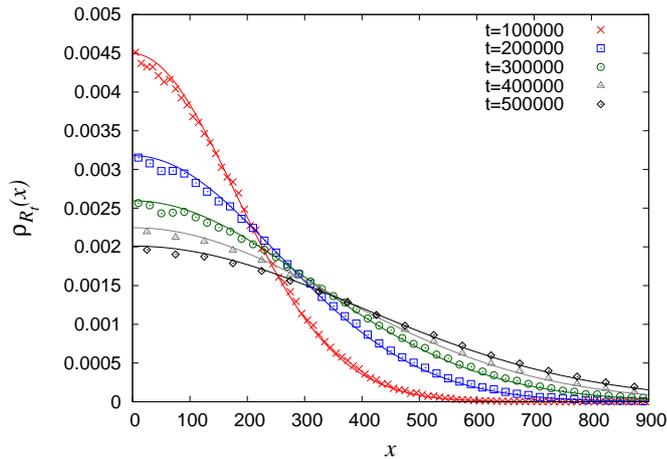}
    \caption{(Color online) Comparison between histograms of the positive displacement performed by the experiment with Eq. (\ref{eq:rhopn}). The diffusion process was generated by Eq. (\ref{eq:chaing}) with $\alpha=0.5$. Each one of the $2.5\times 10^5$ particles uniformly distributed in the interval $[-1/2,1/2]$ is iterated $t$ times. The red crosses, blue squares, green circles, grey triangles and black losangles represent respectively the histograms for times $t=1\times10^5$, $2\times10^5$, $3\times10^5$, $4\times10^5$ and $5\times10^5$. The red, blue, green, grey and black lines (up to down at $x=0$) are Eq. (\ref{eq:rhopn}) calculated for $\beta=2.52$ and respective $t$ already given. The theoretical curves match almost perfectly our data. The number $\beta$ was evaluated by inspection.}
\label{fig:histogram05}
\end{figure}

FIG. \ref{fig:histogram05} shows an example of a comparison between Eq. (\ref{eq:rhopn}) and the respective data for fixed $\alpha$ and $a$ and different values of $t$. The match between them is great. For $10^6$ initial conditions uniformly distributed in $[-1/2,1/2]$, we have chosen times after $t=4\times 10^5$ to build histograms statistically significant. Similar numerical simulations with variations in $a$ and $\alpha$ confirm ADK theorem as well. We also notice that $\beta$ does not depend on time $t$.

Note the reader that finding Mittag-Leffler distributions in both random variables $R_t$ and $L_t$ is quite reasonable. Observing the structure of Geisel-Thomae map, one can see the presence of fixed points in all integers. In studies of Pomeau-Manneville maps, such fixed points are singularities in the invariant measure of the map, which produces small intervals around itself, called laminar regions, where the particle spend most part of the time of its trajectory. In this context, plenty of works has studied that the number of first-passage times $N_t$ to the particle to return to laminar region obeys a Mittag-Leffler distribution \cite{gaspard1988, aizawa2007, venegeroles2012, naze2014}. Thus, as the particle has two senses to return to some laminar region, and the displacements $R_t$ and $L_t$ are proportional respectively to $N_t$ in the right or left senses, it is quite natural to conclude that both variables must obey Mittag-Leffler distributions as well.

\begin{figure}
 \centering
 \includegraphics[scale=0.7]{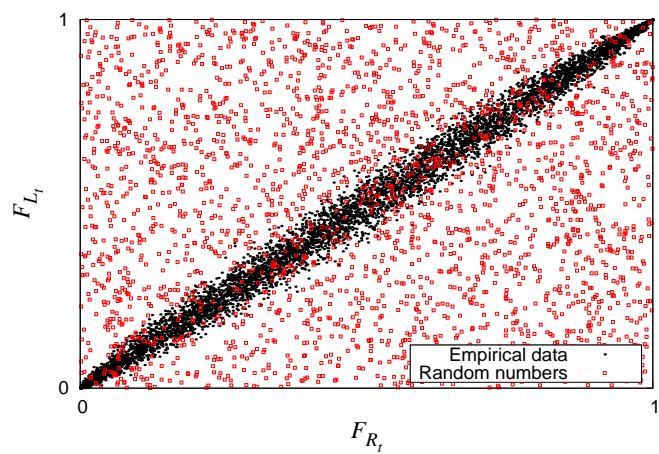}	
\caption{(Color online) Comparison between the scatter plot of pseudo-observations (black full circles) and random numbers (red empty squares) generated for product copula. The disagreement is evident. It was used $\alpha=0.5$. Similar results occur for $\alpha=0.3, 0.4, 0.6$ as well.}
\label{fig:product}
\end{figure}

Before proceeding in finding the copula distribution, we have to guarantee that the random variables $R_t$ and $L_t$ present a non-trivial statistical dependency between them. This is accomplished if the copula of the system is not equal to the product copula
\be
C(u,v) = uv.
\label{eq:independent}
\ee
FIG. \ref{fig:product} shows a comparison between pseudo-observations data and random numbers created using product copula distribution as the generator. 2500 points are used for the former and 2500 points to the latter. The disagreement is evident. Similar results occur for $\alpha=0.3, 0.4, 0.6$ as well. In this manner, to find a non-trivial copula is necessary.

\subsection{Copula distribution: statistical inference}
\label{subsec:copula}

The objective of this section is to find a parametrized copula $C_{\theta}(u,v)$ that better describes the jump distribution PDF $\rho_{\alpha}(x,t)$ of Geisel-Thomae map via statistical inference. In all hypothesis tests performed, it was used software \texttt{R}, version 3.3.2, with the package \texttt{copula}, version 0.999-14 \cite{rdoc-copula}. All the data generated was made with a time of iteration $t=100000$ \footnote{Because of such long time of iteration, it is considered that the initial conditions are mutually independent of each other, condition necessary to apply copula modeling \cite{nelsen2006}.} and the analyzed cases were $\alpha=0.3, 0.4, 0.5, 0.6$. \texttt{R} offers six possibilities to analyze: Clayton, Frank, Gumbel-Hougaard, normal, Plackett and $t$-Student (hereafter taking the degrees of freedom $\nu=4$) families (see the mathematical expressions at the end of this subsection). The specific hypothesis test used was the Cram\'er-Von Mises test, using the multiplier bootstrap as a method to estimate the $p$-values \cite{degroot2011}. It was used the maximum pseudolikelihood (MPL) as a method to estimate the parameters $\theta$ that characterize the copula family \footnote{Under various possibilities to execute such test, the options chosen here seem the most appropriate. Multiplier bootstrap is more computationally efficient than the parametric bootstrap \cite{kojadinovic2011}, and the Anderson-Darling test, which is an option to the Cram\'er-Von Mises test, is on debate in the scientific community about the values of parameters that compound it \cite{genest2012}, and we decided not to analyze our data under such perspective. Lastly, previous tests with inversion of Kendall's $\tau$ and inversion of Pearson's $\rho$ as estimative methods of the parameters did not give any new insight.}. 

\begin{table}[!ht]
\begin{center}
\begin{tabular}{|c||c|c|}
\hline
\hline
\multicolumn{3}{|c|}{\textbf{Hypothesis test}}\\
\hline
Copula &$\theta$ &$p$-value($\%$) \\
\hline
Clayton &10.55 &$10^{-5}$\\
\hline
Frank &42.82 &0.023\\
\hline
Gumbel-Hougaard &9.19 &0.697\\
\hline
Normal &0.967 &$10^{-5}$\\
\hline
Plackett &442.64 &0.033\\
\hline
$t$-Student &0.988 &0.661\\
\hline
\hline
\end{tabular}
\caption{Table presenting the outcome from the command line \texttt{gofCopula} for the Clayton, Frank, Gumbel-Hougaard, normal, Plackett and $t$-Student cases taken under null hypothesis for $\alpha=0.5$. $\theta$ is the value of the parameter estimated by MPL method and the $p$-values were evaluated by multiplier bootstrap, with the exception of Frank case, where parametric bootstrap was performed. The sample size $n=750$ and the number of bootstraps repetitions $m=50000$ were used. All the cases were rejected. Same results for $\alpha=0.3, 0.4, 0.6$.}
\label{table:1}
\end{center}
\end{table}

\begin{figure*}
\centering
 \includegraphics[scale=0.46]{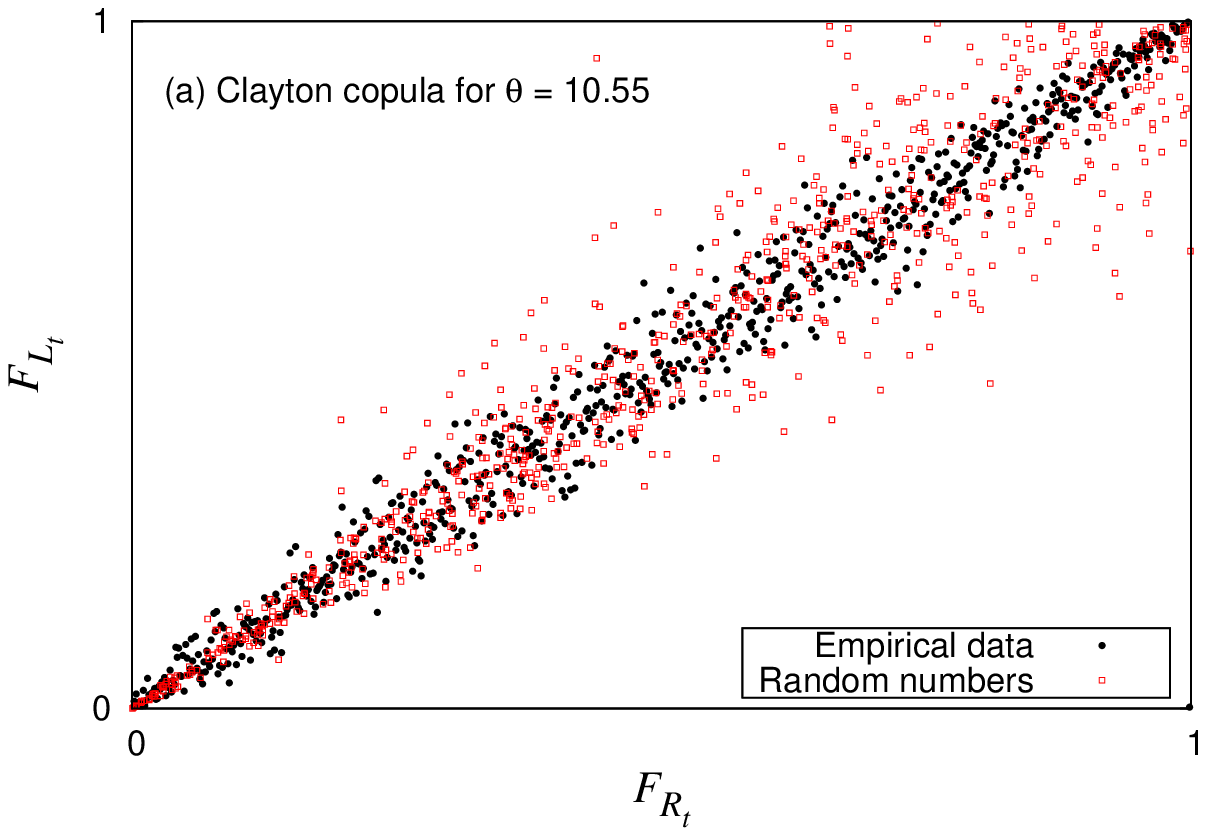}
 \includegraphics[scale=0.46]{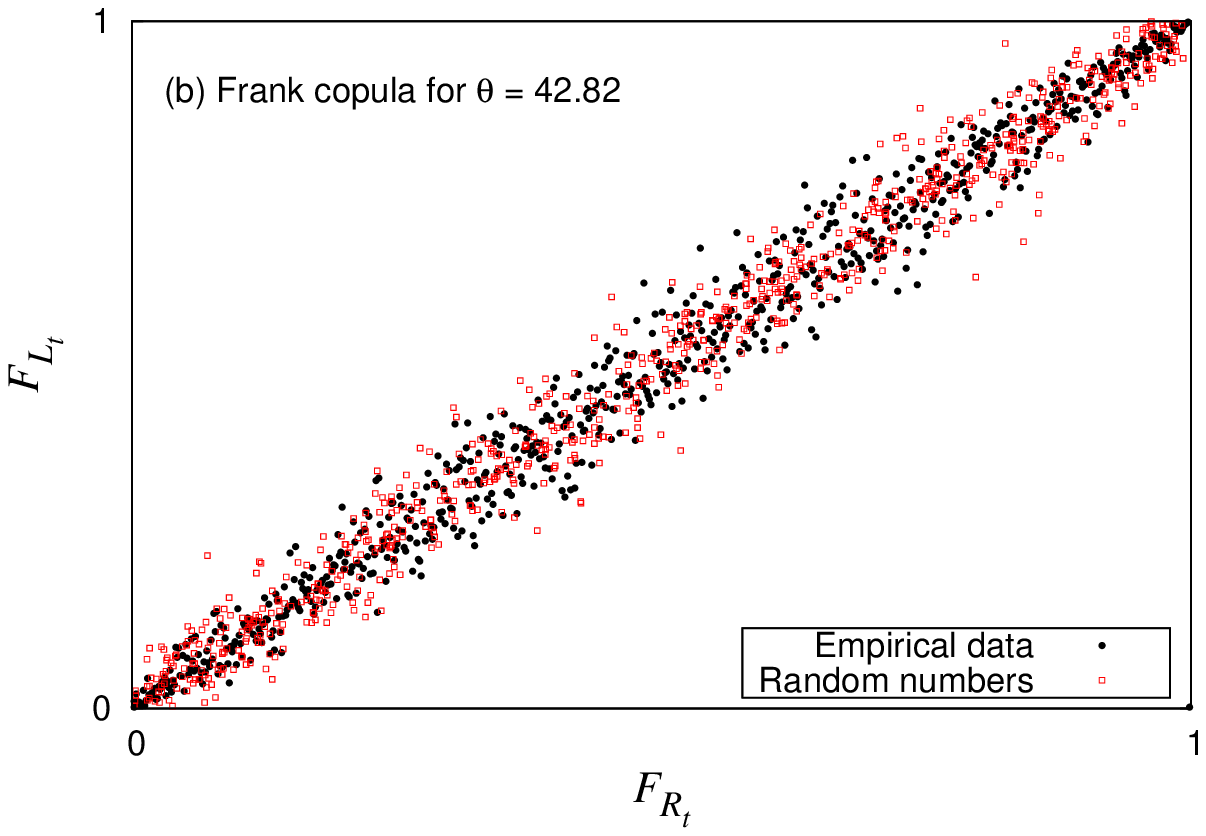}
 \includegraphics[scale=0.46]{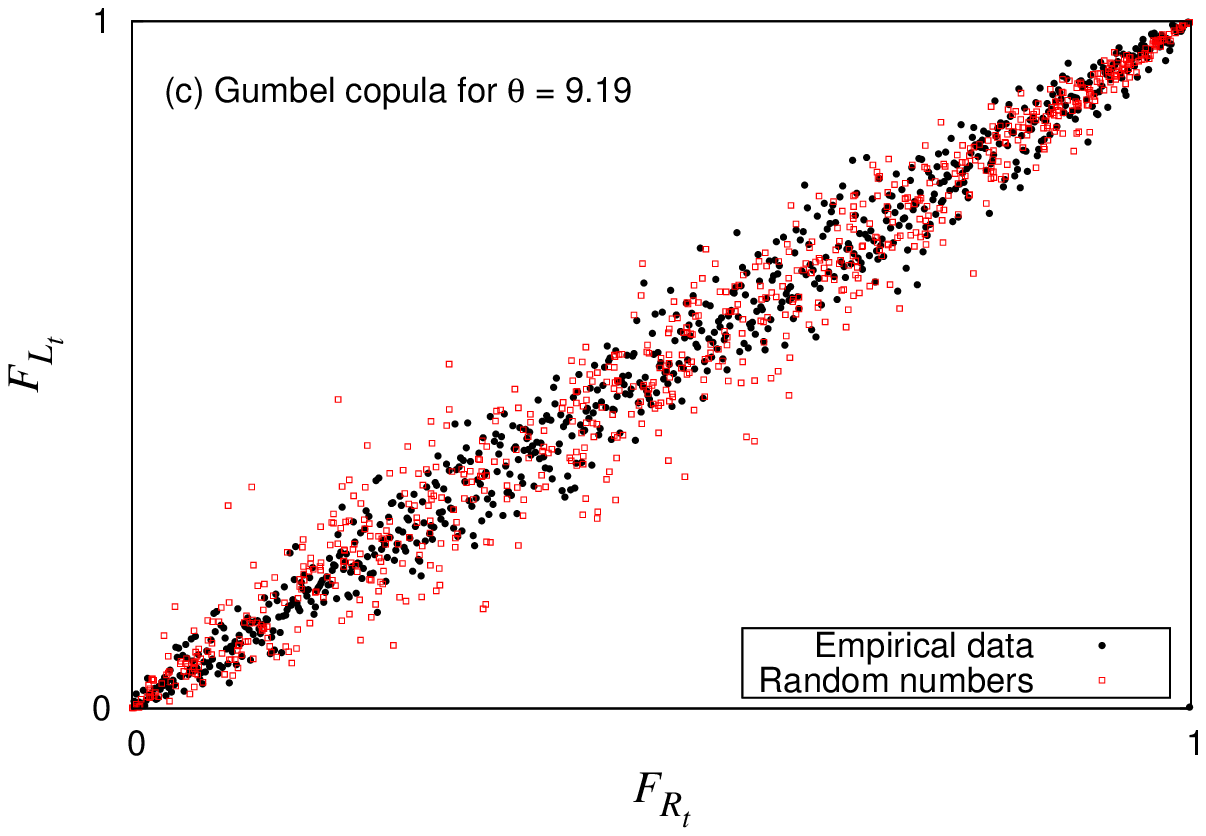}
 \includegraphics[scale=0.46]{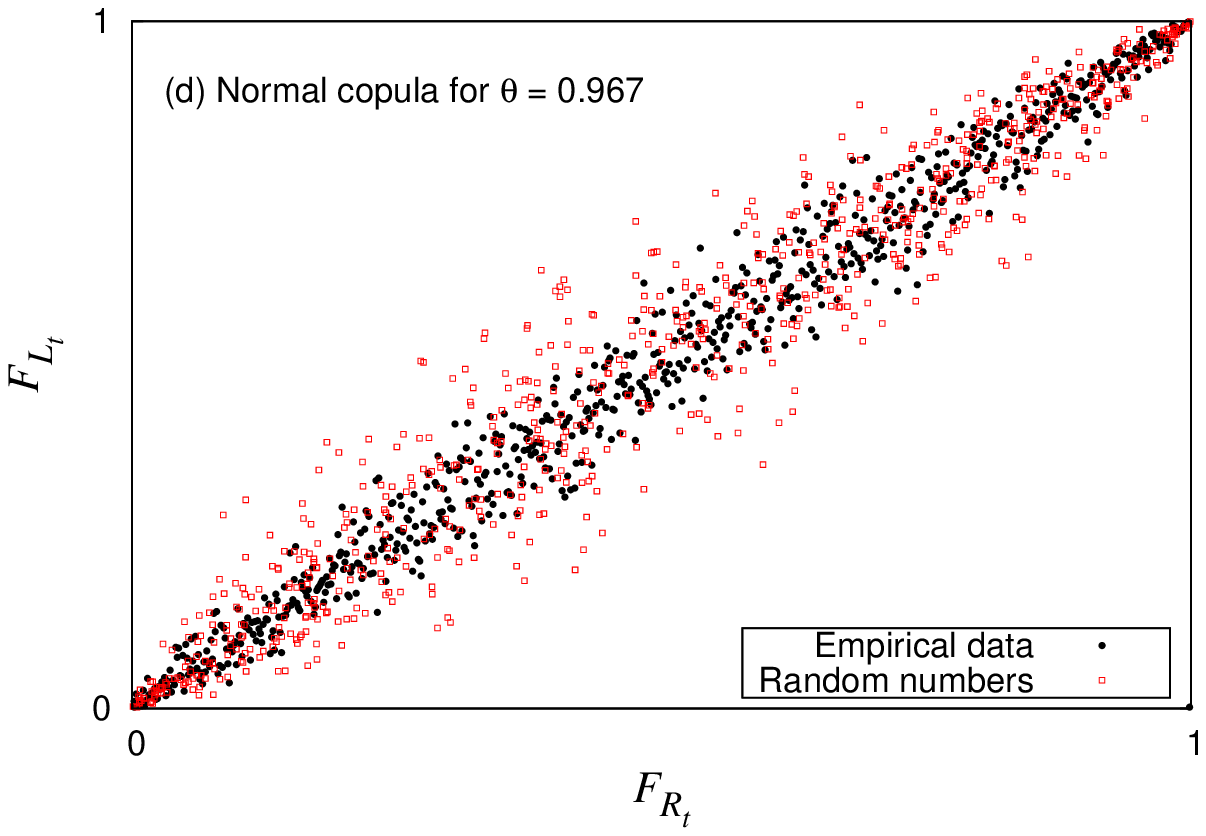}
 \includegraphics[scale=0.46]{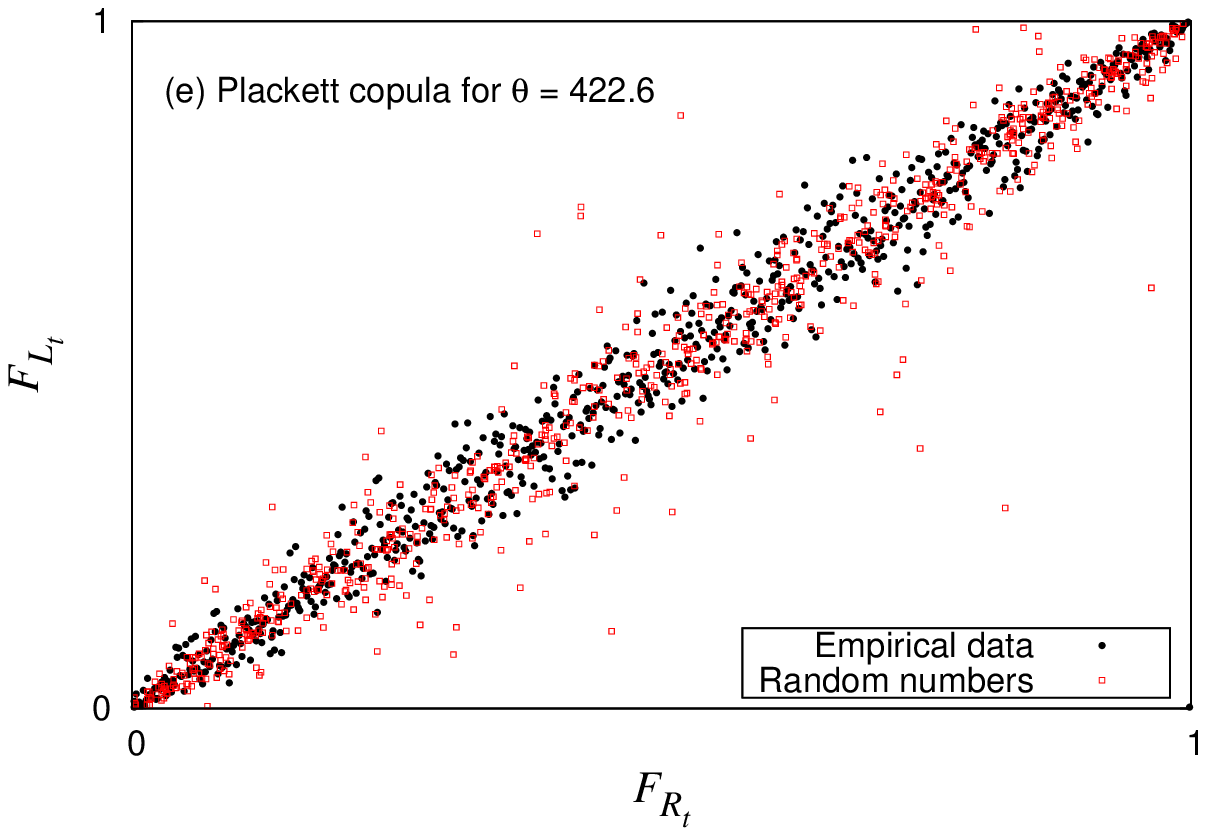}
 \includegraphics[scale=0.46]{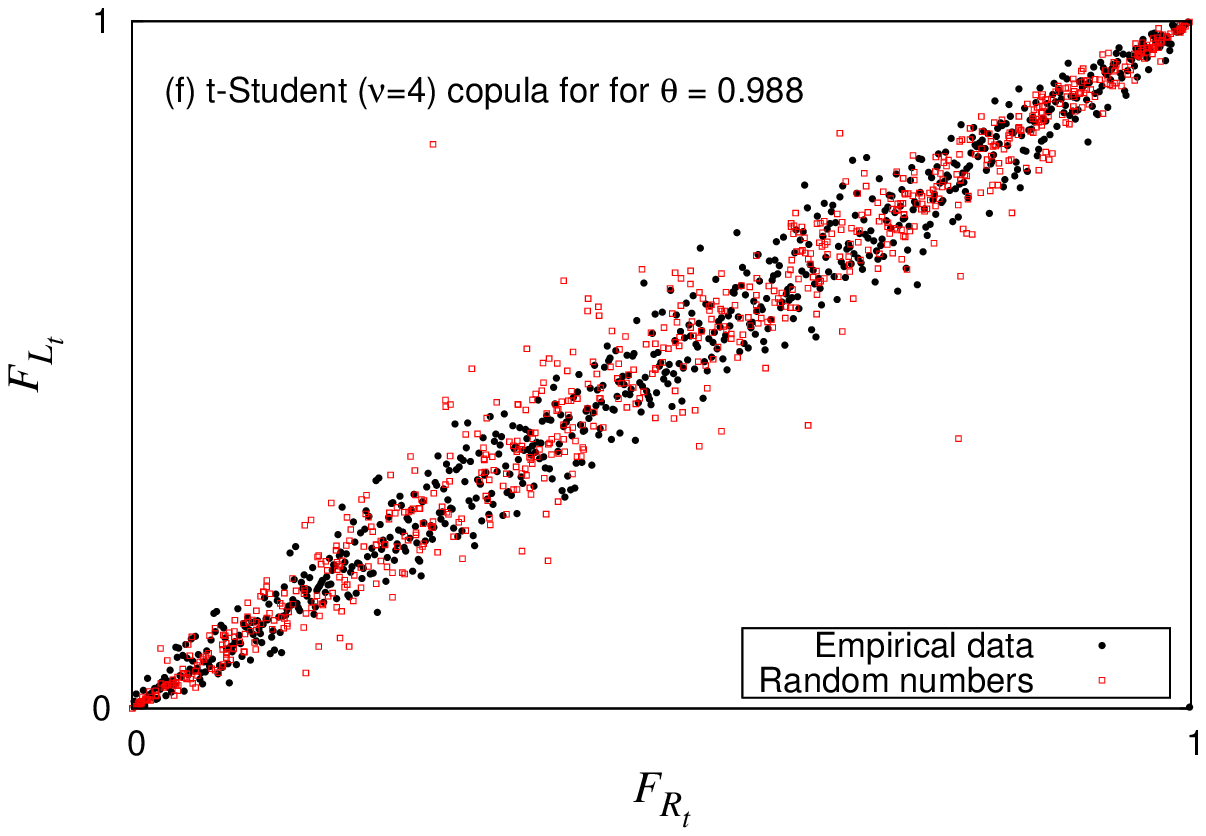}	
\caption{(Color online) Comparison between the scatter plot of pseudo-observation (black full circles) and random numbers (red empty squares) generated for (a) Clayton, (b) Frank, (c) Gumbel-Hougaard, (d) normal, (e) Plackett and (f) $t$-Student copula families for $\alpha=0.5$. The parameters were estimated by maximum pseudolikelihood method. For each figure, it was used 750 points. All the graphics present some disagreements between the plots, which corroborates the results presented in TABLE I.}
\label{fig:copulas}
\end{figure*}

TABLE \ref{table:1} provide the outcomes from the hypothesis tests performed taking Clayton, Frank, Gumbel-Hougaard, normal, Plackett and t-Student copulas under null hypothesis for $\alpha=0.5$. It was used the command \texttt{gofCopula}, a sample size $n=750$, a number of bootstraps repetitions $m=50000$ and a significance level of $1\%$. All the cases are rejected under the null hypothesis at the significance level of $1\%$. The same results occur for $\alpha=0.3,0.4, 0.6$. FIG. \ref{fig:copulas} illustrates the results obtained in TABLE \ref{table:1}. It depicts a comparison between the scatter plot for pseudo-observations and random numbers created for (a) Clayton, (b) Frank, (c) Gumbel-Hougaard, (d) normal, (e) Plackett and (f) $t$-Student copula distributions used as generators. On each graphic, 750 points were used in the scatter plot for pseudo-observations and 750 points for the random numbers plot. All the random numbers plots, in some way or another, do not fit the pseudo-observations data. Describing respectively the figures of the worst and best case scenario from TABLE \ref{table:1}, the Clayton copula fits the region near the left tail, but disagrees completely near the right one, and Gumbel-Hougaard copula, although fits almost perfectly on the tails, is spread in the remaining part of the plot.

Although the result could be possibly better with more options of copula families at our disposal, note the reader that hypothesis test must be the first technique used to determine the copula distribution, because it says whether the analyzed copula family is suitable or not. In the case which is not, we have to use other methods to find a better approximation. In our case, we evaluate the best model for the six families observing their agreements with the jumps distribution. Thus, we perform $\chi^2$ tests \cite{degroot2011} considering the points of the histogram generated by Geisel-Thomae map as the input of the theoretical part and those points evaluated by Eq. (\ref{eq:difffinal}) as the input of the experimental one. It was used $n=250000$, $\beta=2.00, 2.23, 2.52, 3.00$ (verified by inspection) respectively for the cases $\alpha=0.3, 0.4, 0.5, 0.6$, where we take the number of points for each $\chi^2$ test respectively 30, 60, 100, 140 points. The parameter estimation was evaluated by the command \texttt{fitCopula} and the histograms were constructed with unitary bins, meaning that we only observe the dynamics coarse behavior \cite{zumofen1993}.

\begin{table}
\centering
\begin{tabular}{|c||c|c|c|c|c|c|c|c|c|c|c|c|}
\hline
\hline
\multicolumn{13}{|c|}{Evaluations of $\theta$ and $\chi^2$}\\ \cline{2-13}
\hline
Copula & \multicolumn{2}{|c|}{Clayton} & \multicolumn{2}{|c|}{Frank} & \multicolumn{2}{|c|}{\bf Gumbel-Hougaard} & \multicolumn{2}{|c|}{Normal} & \multicolumn{2}{|c|}{Plackett} & \multicolumn{2}{|c|}{$t$-Student}\\
\hline
$\alpha$ &$\theta$ &$\chi^2$ &$\theta$ &$\chi^2$ &${\bf \theta}$ &${\bf \chi^2}$ &$\theta$ &$\chi^2$ &$\theta$ &$\chi^2$ &$\theta$ &$\chi^2$ \\
\hline
0.3 &3.57 &1.28 &17.14 &2.81 &{\bf 4.41} &{\bf 0.002} &0.942 &0.021 &70.49 &0.057 &0.933 &0.023\\
\hline
0.4 &6.39 &2.51 &28.65 &4.94 &{\bf 6.82} &{\bf 0.004} &0.976 &0.009 &185.5 &0.059 &0.973 &0.014\\
\hline
0.5 &10.97 &8.74 &44.97 &8.6 &{\bf10.14} &{\bf0.003} &0.989 &0.012 &445.57 &0.055 &0.988 &0.012\\
\hline
0.6 &17.54 &8.23 &64.61 &13.96 &{\bf 13.84} &{\bf 0.006} &0.995 &0.011 &913.16 &0.042 &0.994 &0.008\\
\hline
\hline
\end{tabular}
\caption{Table presenting the values of $\theta$ and $\chi^2$ for Clayton, Frank, Gumbel-Hougaard, normal, Plackett and t-Student families for the cases $\alpha=0.3$, 0,4, 0.5, 0.6. Gumbel-Hougaard copulas present the better agreements.}
\label{table:2}
\end{table}

\begin{figure}
 \centering
 \includegraphics[scale=0.46]{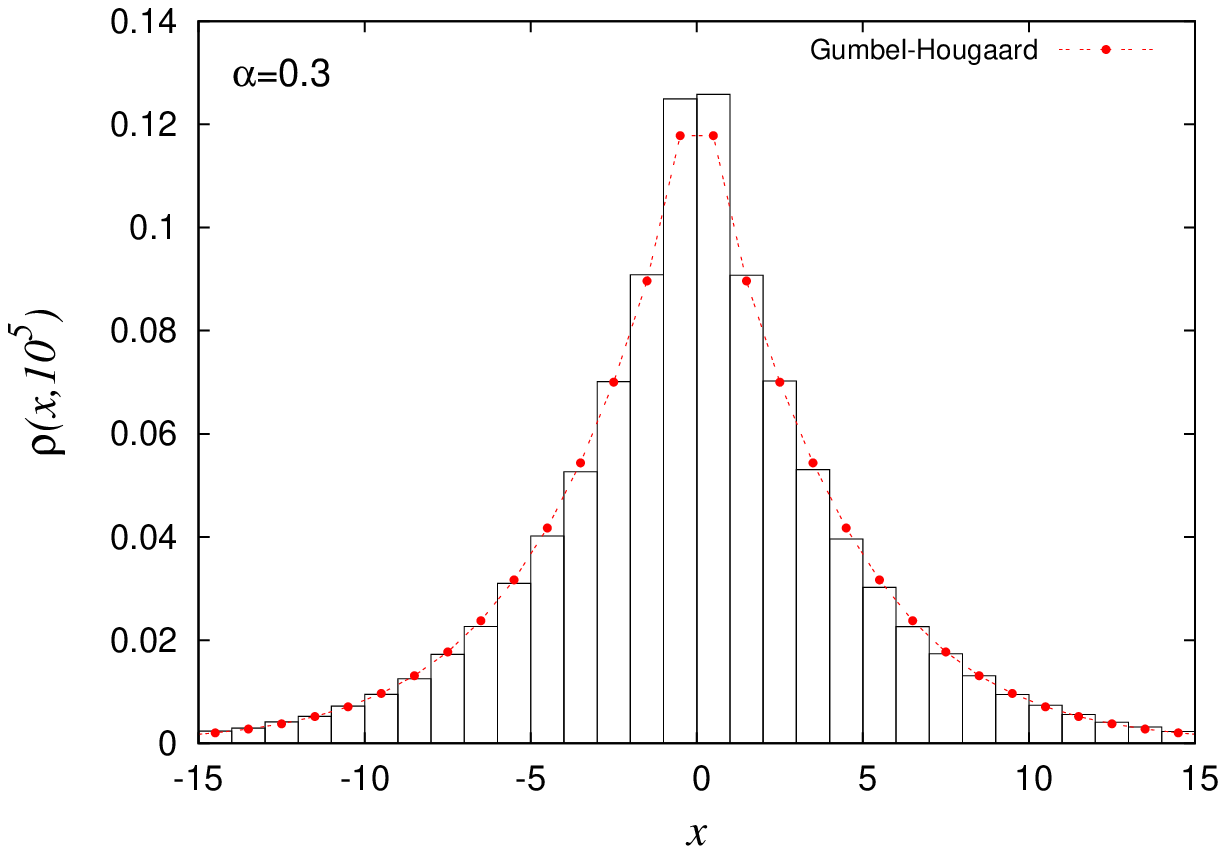}
 \includegraphics[scale=0.46]{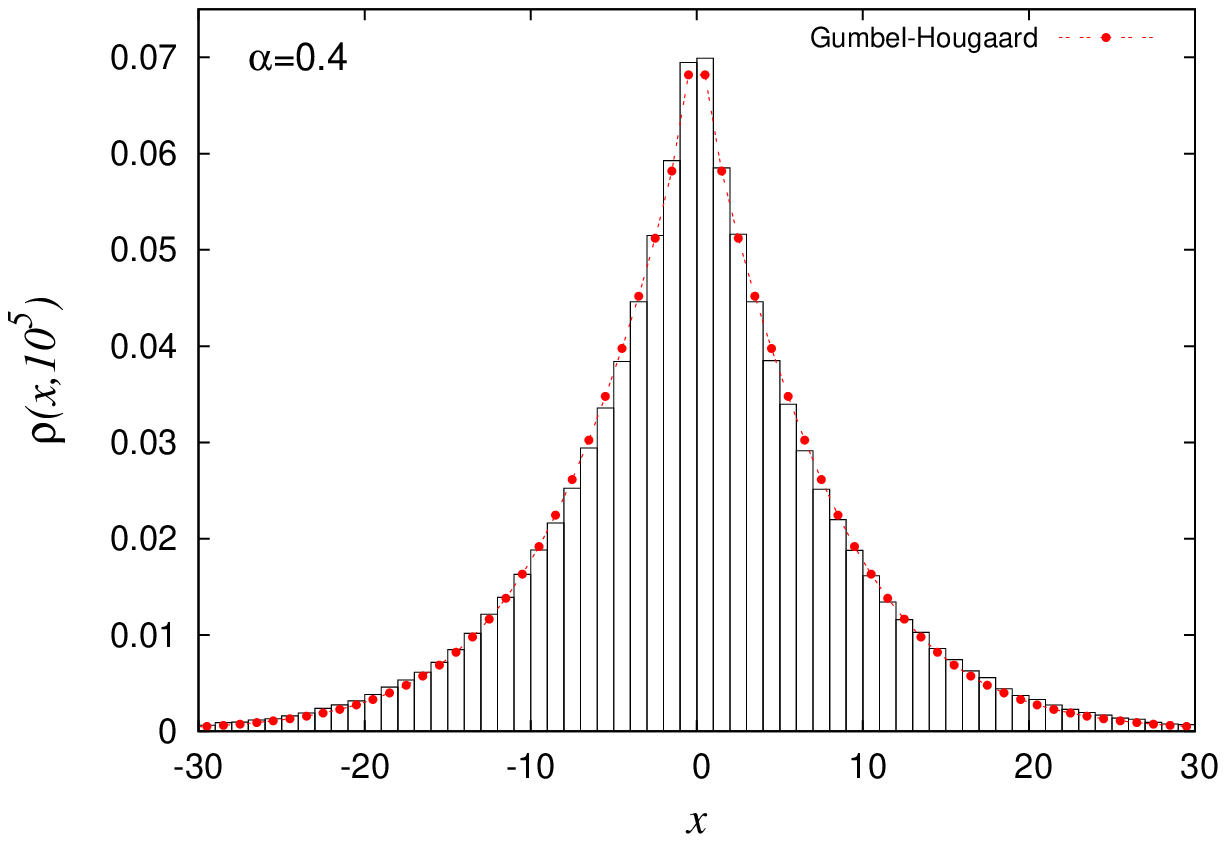}
 \includegraphics[scale=0.46]{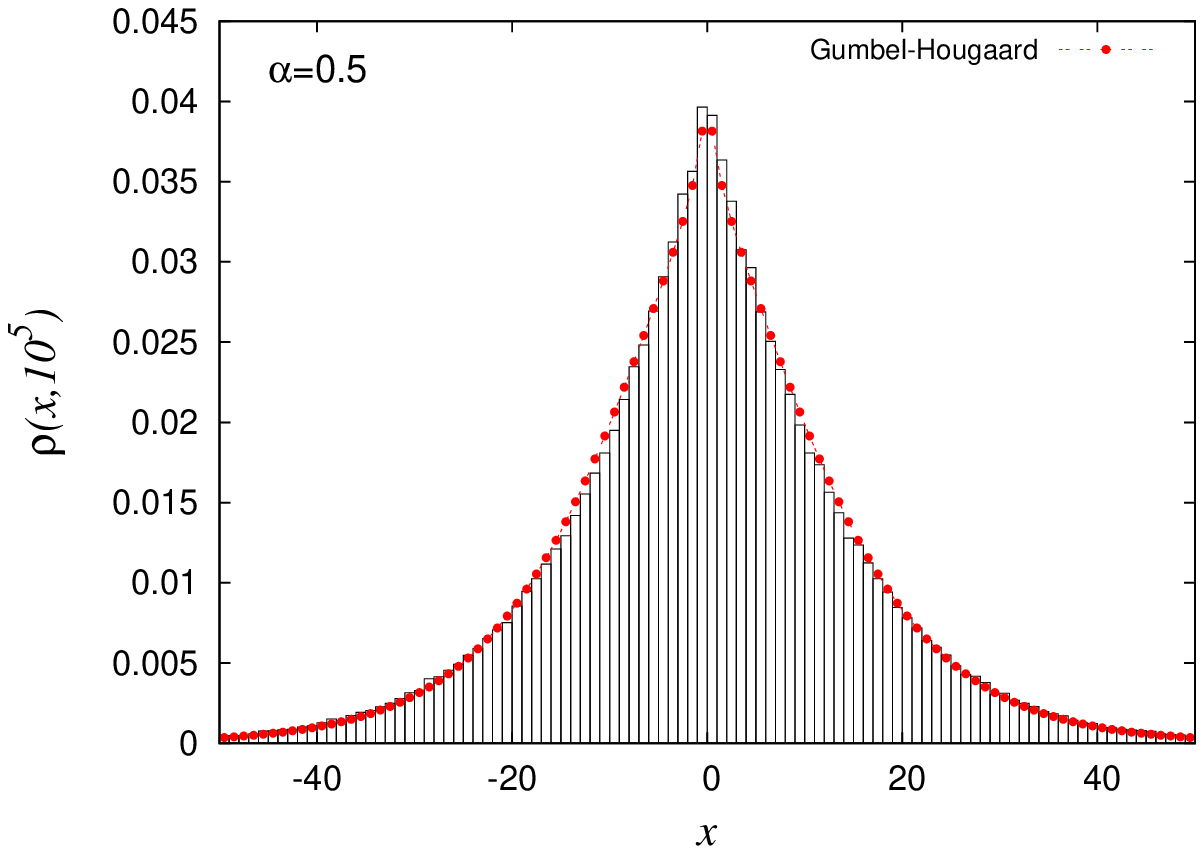}
 \includegraphics[scale=0.46]{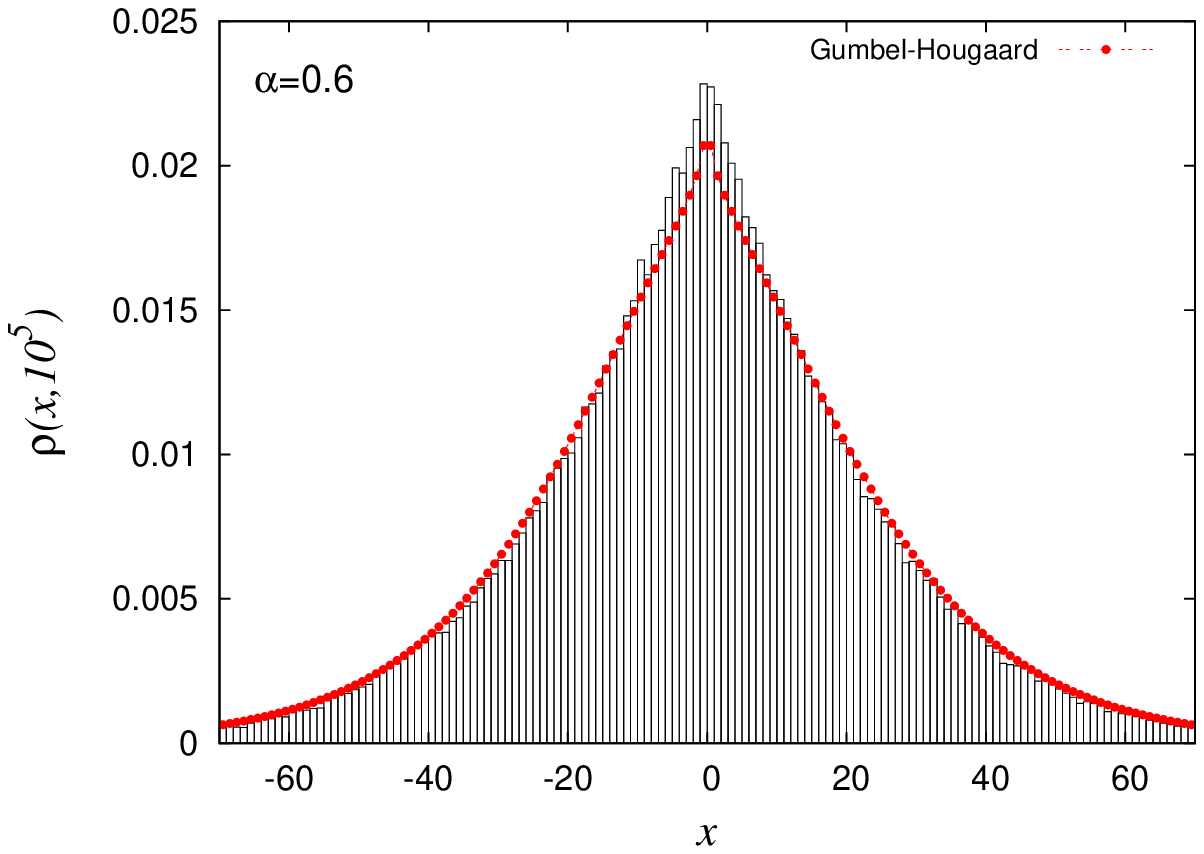}
\caption{(Color online) Comparisons between the PDFs generated by Gumbel-Hougaard case (red circles) with the histogram produced by Geisel-Thomae map for $\alpha=0.3, 0.4, 0.5, 0.6$. The parameter estimation was done by MPL method with $n=250000$. For all cases, the better agreement occurs with the tails of the histogram.}
\label{fig:comp_MPL}
\end{figure}

TABLE \ref{table:2} presents the results. For the cases $\alpha=0.3, 0.4, 0.5 , 0.6$, Gumbel-Hougaard copula presents respectively $\chi^2 = 0.002, 0.004, 0.003, 0.006$, lesser than the values of the other five families analyzed, being therefore the best copula approximation. In particular, FIG. \ref{fig:comp_MPL} shows that the PDFs generated by such copula have an excellent agreement with the tails of the histograms, which is a quite interesting property, because the tail behavior of the jumps distribution is a manner of distinguishing anomalous processes \cite{zumofen1993}. Therefore, the kind of subdiffusion treated here can be characterized by a Gumbel-Hougaard copula coupling Mittag-Leffler distributions -- a feature of its dynamics. The fact that such copula, which is an extreme value one, is describing a tail distribution is not fully understood yet and it will be subject for future research. We also remark that other parametrized families graphics do not present any particular region of agreement.

\begin{figure}
\centering
 \includegraphics[scale=0.7]{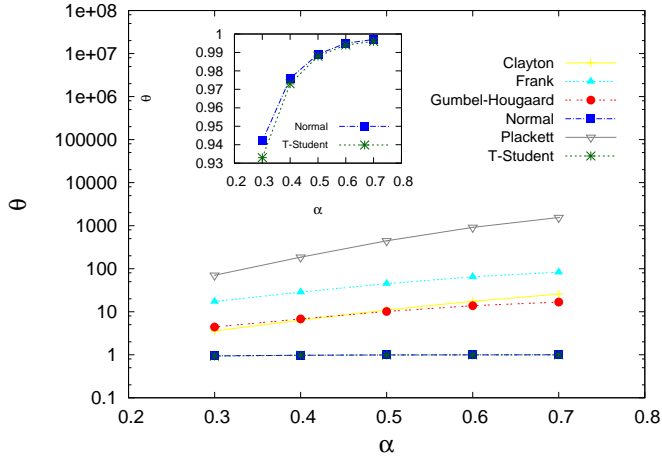}
\caption{(Color online) Dependency between the estimated $\theta$ for Clayton (yellow crosses), Frank (light blue triangle), Gumbel-Hougaard (red circles), normal (blue squares), Plackett (gray inverted triangle) and $t$-Student (green asterisk) and $\alpha$. The inset graphic is a zoom of the plots of Normal and $t$-Student family. All parameters grows as $\alpha$ is increased. For the point of view of statistical dependency across $R_t$ and $L_t$, they become more dependent as $\alpha$ increases.}
\label{fig:dependency}
\end{figure}

FIG. \ref{fig:dependency} shows the graphic between the estimated parameter $\theta$ for each copula family and the parameter $\alpha$. For all the cases, the parameters are proportional to $\alpha$. In this way, as the statistical dependency between the random variables increases with the parameters (see the mathematical expressions in the end of this subsection), the same occurs as $\alpha$ is increased. Such result is consistent with the dynamics of our system. As $\alpha\rightarrow 0$, the branches of Geisel-Thomae map approach to the diagonal, which means, for the dynamical point of view, that the system stays near the neutral points for much time than before. In this way, if the system begins in the right branch of any cell map, most part of the time it will jump only to the right sense. Then the passage for regions where the system would jump to the left sense is almost negligible. $R_t$ is practically unaffected by $L_t$ and therefore they are statistically independent in the limit $\alpha\rightarrow 0$. The same idea is valid considering initial conditions beginning in the left cell branches. 

Finally, for completeness, the analyzed copula families are enumerated below.

\begin{enumerate}

{\it \item  Clayton copula:}
\be
C_{\theta}(u,v) = [\text{max}\{u^{-\theta}+v^{-\theta}-1,0\}]^{-\frac{1}{\theta}},
\ee
where $\theta\in[-1,\infty)\backslash \{0\}$. Independency occurs in the limit $\theta\rightarrow 0$.

\vspace{0.2cm}

{\it \item Frank copula:}
\be
C_{\theta}(u,v) = -\frac{1}{\theta}\log{\left[1+\frac{(e^{-\theta u}-1)(e^{-\theta v}-1)}{e^{-\theta}-1}   \right]},
\ee
where $\theta\in\R \backslash \{0\}$. Independency occurs in the limit $\theta\rightarrow 0$.

\vspace{0.2cm}

{\it \item Gumbel-Hougaard copula:}
\be
C_{\theta}(u,v) = \exp{\left[-\left((-\text{ln} u)^\theta+(-\text{ln} v)^\theta\right)^{\frac{1}{\theta}}\right]},
\ee
where $\theta\ge 1$. Independency occurs at $\theta=1$.

\vspace{0.2cm}

{\it \item Normal copula:}
\be
C_{\rho}(u,v) = \frac{1}{\sqrt{2\pi(1-\rho^2)}}\int\displaylimits_{-\infty}^{F^{-1}(v)} \int\displaylimits_{-\infty}^{F^{-1}(u)} e^{\left(\footnotesize{-\frac{x^2+y^2-2\rho x y}{2(1-\rho^2)}} \right)}dxdy, 
\ee
where $\rho$ is the Pearson's correlation coefficient and $F^{-1}$ is the inverse CDF of univariate normal distribution. Independency occurs to $\rho=0$.

\vspace{0.2cm}

{\it \item Plackett copula:}
\be
C_{\theta}(u,v) = \frac{1+(\theta-1)(u+v)-\sqrt{{[1+(\theta-1)(u+v)]}^2-4\theta(\theta-1)uv}}{2(\theta-1)}, 
\ee
where $\theta>0$. Independency occurs at $\theta=1$.

\vspace{0.2cm}

{\it \item $t$-Student ($\nu=4$) copula:}
\be
C_{\rho}(u,v) =  \frac{\Gamma(3)}{4\Gamma(2)\sqrt{\pi(1-\rho^2)}} \int\displaylimits_{-\infty}^{t^{-1}_{4}(v)} \int\displaylimits_{-\infty}^{t^{-1}_{4}(u)}{\left(1+\frac{x^2+y^2-2\rho xy}{4(1-\rho^2)} \right)}^{-3}dxdy, 
\ee
where $\rho$ is the Pearson's correlation coefficient, $t_{4}^{-1}$ is the inverse CDF of univariate $t$-Student distribution for $\nu=4$. Independency occurs at $\rho=0$.
\end{enumerate}

\subsection{Copula and joint distributions: exact results}

We claim that the joint distribution $J_t$ can be calculated exactly once we know the jumps distribution $\rho(x,t)$. We observe first that $\rho(x,t)$ can be expressed as
\be
\rho(x,t) = \int_{0}^{\infty} \frac{\partial}{\partial z}\left(-\frac{\rho(|x|+z,t)\rho(z,t)}{\rho(0,t)}\right)dz.
\label{eq:copulaalpha}
\ee
Then we equal such expression with Eq. (\ref{eq:difffinal}) and isolate the copula PDF $c_t$. Passing into variables $(u,v)$, we have
\be
c_{t}(u,v) = \left. -\frac{\rho(y,t)\partial_x \rho(x,t)+\rho(x,t)\partial_y \rho(y,t)}{\rho(0,t)\rho_{R_t}(x)\rho_{L_t}(y)}\right|_{\substack{x=F_{R_t}^{-1}(u) \\ y=F_{L_t}^{-1}(v)}},
\label{eq:physcopula}
\ee
where $F_{R_t}^{-1}$ and $F_{L_t}^{-1}$ are respectively the inverse CDF of $R_t$ and $L_t$. In the case of Geisel-Thomae map, $\rho(x)$ will be Eq. (\ref{eq:rrrrho}), whose partial derivative $\partial_x \rho_{\alpha}(x,t)$ is easily computable. Furthermore, based on the approach of \cite{saa2011}, we can express
\be
\rho_{R_t}(x) = \rho_{L_t}(x) = \frac{1}{\alpha}x^{-(1+1/\alpha)}g_{\alpha}(x^{-1/\alpha}),
\ee
where $g_{\alpha}(x)$ is one-sided L\'evy PDF , which in turn can be expressed analytically by Mikusinski's integral representation. The inverse CDF of $R_t$ and $L_t$ can be in principle calculable by this same representation. Indeed, we have
\be
F_{R_t}^{-1}(x) = {\log{(f(x))}}^{1-\alpha}, 
\ee
where $f(x)$ is the solution of the following integral equation
\be
\int_0^\pi f(x)^{w(\phi)}d\phi = \pi(1-x),
\ee
with
\be
w(\phi) = \frac{\sin{(1-\alpha)}\phi}{\sin{\phi}}{\left(\frac{\sin{\alpha\phi}}{\sin{\phi}}\right)}^{\alpha/(1-\alpha)}.
\ee
At this point is easy to see the necessity of inference tests to find an approximate copula, since the exact is hard to be analytically computatable. We remark also that Eq. (\ref{eq:physcopula}) is symmetrical, that is, $c_t(u,v)=c_t(v,u)$, as we have verified in the cumulative case in FIG. \ref{fig:product} and FIG. \ref{fig:copulas} . Finally, using Sklar theorem in its PDF version in Eq. (\ref{eq:physcopula}), the jointly PDF of the variables $R_t$ and $L_t$ will be given by
\be
j_t(x,y) = -\frac{\rho(y,t)\partial_x \rho(x,t)+\rho(x,t)\partial_y \rho(y,t)}{\rho(0,t)}.
\ee
In the case of Geisel-Thomae map, where the jumps distribution $\rho_{\alpha}(x,t)$ could be determined by CTRW method, we have
\be
j_t(x,y) = -\frac{\rho_{\alpha}(y,t)\partial_x \rho_{\alpha}(x,t)+\rho_{\alpha}(x,t)\partial_y \rho_{\alpha}(y,t)}{\rho_{\alpha}(0,t)}.
\ee  
In other words: the joint distribution of two random variables is determined if one knows the distribution of their difference.

\section{Final remarks}
\label{sec:4}

We presented in this work the connection between the statistical distributions of weakly chaotic dynamics and deterministic subdiffusion. Considering Geisel-Thomae map, such relation was established by Sklar theorem, where the jumps distribution was decoupled into Mittag-Leffler distributions and a Gumbel-Hougaard copula for different subdiffusion parameters $\alpha$. We presented also a method to calculate the exact copula distribution of the system, although under the condition that the statistical distributions of weakly chaotic dynamics and deterministic subdiffusion are known. In the end, we observed that the copula parameters, which measure the statistical dependency between the marginal distributions, is proportional to the subdiffusion parameter $\alpha$, being consistent therefore with the dynamics of the system.

If in the past copula modeling was a technique hard to be put on practice, because it was practically impossible to analyze a considerable amount of data of an statistical experiment, at the present moment studies can be performed easily by the recent development of software packages. Besides that, new techniques in copula modeling are in constant development and the community behind all these advances increasingly grows. In this manner, physical phenomena in which statistical dependency is a fundamental subject for their understanding can be investigated by this new perspective. Some examples that could be addressed in this manner are the hypothesis of molecular chaos in Boltzmann equation or the map families defined by Gaspard and Wang in \cite{gaspard1988}.    

Finally, it is important to stress that copulas are not just empty mathematical functions waiting to describe problems in a redundant way, but carries important properties that help to understand the studied phenomena. For example, Gumbel-Hougaard copula has its roots in the extreme value theory, which means that it is suitable to model dependency between extreme events, such as a possible flood in the city by the water level of the rivers that surround it \cite{gumbel1964}. On the other hand, knowing that Gumbel-Hougaard copula has appeared naturally in a problem, suggests that the dependency across the variables obeys at some extent the extreme value theory. In this manner, as we have obtained that Gumbel-Hougaard copula describes the tail of jumps distribution, this probably implies some deeper connection between extreme value copulas and tail distributions. This aspect will be subject for future research.

\section*{Acknowledgements}
I thank Marcus V. S. Bonan\c{c}a, Alberto Saa and Roberto Venegeroles for reading the manuscript; A. Suzuki and M. L. Viola for discussions about copula modeling. This work was financed by CNPq and CAPES.

\bibliographystyle{spmpsci}

\end{document}